\documentclass[aps,superscriptaddress,nofootinbib,eqsecnum,prd,notitlepage,twocolumn]{revtex4-1} 

\pdfoutput=1

\usepackage{amsfonts}
\usepackage{amsmath}
\usepackage{amssymb}
\usepackage{graphicx,color}
\usepackage{float}
\usepackage{hyperref}
\usepackage{subfigure}
\usepackage{dcolumn}
\usepackage{soul}
\usepackage{ulem}
\usepackage{verbatim}

\begin{document}

\title{Dirac-Born-Infeld warm inflation realization in the strong
  dissipation regime}

\author{Meysam Motaharfar}
\email{mmotaharfar2000@gmail.com}
\affiliation{Department of Physics and Astronomy, Louisiana State University,
  Baton Rouge, LA 70803, USA}

\author{Rudnei O. Ramos} \email{rudnei@uerj.br}
\affiliation{Departamento de Fisica Teorica, Universidade do Estado do
  Rio de Janeiro, 20550-013 Rio de Janeiro, RJ, Brazil }

\begin{abstract}

We consider warm inflation with a Dirac-Born-Infeld (DBI) kinetic term
in which both the nonequilibrium dissipative particle production and
the sound speed parameter slow the motion of the inflaton field. We
find that a low sound speed parameter removes, or at least strongly
suppresses, the growing function appearing in the scalar of curvature
power spectrum of warm inflation, which appears due to the temperature
dependence in  the dissipation coefficient. As a consequence of that,
a low sound speed helps to push warm inflation into the strong
dissipation regime, which is an attractive regime from a model
building and phenomenological perspective. In turn, the strong
dissipation regime of warm inflation softens the microscopic
theoretical constraints on cold DBI inflation. The present findings,
along with the recent results from swampland criteria, give a strong
hint that warm inflation may consistently be embedded into string
theory. 

\end{abstract}

\maketitle

\section{Introduction}

Warm inflation
(WI)~\cite{Berera:1995wh,Berera:1995ie,Berera:1998px} is an
alternative dynamical realization for conventional cold inflation
(CI)~\cite{Guth:1980zm,Sato:1981ds,Albrecht:1982wi,Linde:1981mu,Linde:1983gd}
during which the inflaton field dissipates its vacuum energy into an
ambient radiation bath, thus eliminating the necessity of a
postinflationary reheating process~\cite{Berera:1996fm}  and leading
into different possibilities for a graceful exit
mechanism~\cite{Das:2020lut}. Such a nonequilibrium dissipative
particle production process acts to slow the motion of the inflaton
field, allowing the embedding of steeper potentials in the WI context
and helping to solve, for example, the  so-called
$\eta$-problem~\cite{Berera:1999ws,BasteroGil:2009ec,Bastero-Gil:2019gao}. In
WI the thermal fluctuations play the dominant role in the formation of
the seeds for the large scale structure (LSS) by producing a quasi
scale-invariant spectrum of primordial curvature perturbations. At the
same time, the shape of the produced spectrum depends on the field
content of the model, leading to signatures able to make it
distinguishable from the one produced in
CI~\cite{Bastero-Gil:2014raa,Moss:2011qc,Moss:2007cv,Bastero-Gil:2014jsa,Ramos:2013nsa,BasteroGil:2011xd,Graham:2009bf,Hall:2003zp,Bastero-Gil:2014oga}.
{}Furthermore, the rich dynamics of WI allows it to address/alleviate
some of long-lasting problems related to the (post-)inflationary phase
in the CI
scenarios~\cite{Bartrum:2012tg,Sanchez:2010vj,Motaharfar:2021gwi,Bastero-Gil:2016mrl,BasteroGil:2011cx,Rosa:2019jci,Dimopoulos:2019gpz,Levy:2020zfo,Rosa:2018iff,Lima:2019yyv,Sa:2020fvn,Berghaus:2019cls}
(see also
Refs.~\cite{Dymnikova:1998ps,Dymnikova:2001ga,Dymnikova:2001jy} for
related work).

Despite its tremendous success, earlier particle physics realizations
of WI were confronted with two important difficulties. The first one
was that achieving a thermal radiation bath during inflation can
result in potentially large thermal corrections to the inflaton's
potential, thus, hindering the slow-roll dynamics. Earlier WI model
building constructions circumventing this problem made use of models
with large field
multiplicities~\cite{Moss:2006gt,BasteroGil:2010pb,BasteroGil:2012cm,Bartrum:2013fia},
making them technically unappealing. This issue was later resolved
with the introduction of a new class of WI model building realization
able to sustain a nearly-thermal bath, yet with a small number of
field species~\cite{Bastero-Gil:2016qru}. These type of models were
dubbed ``warm little inflaton (WLI)" models. The second difficulty was
that the backreaction of the thermal radiation bath on the inflaton
power spectrum due to a temperature dependent dissipation coefficient
leads to the appearance of growing/decreasing modes in the scalar of
curvature power spectrum~\cite{Graham:2009bf}. As a consequence of
this, consistency with the observations could only be achieved in weak
dissipation regimes of WI, thus preventing WI from going into a strong
dissipation regime. However, the strong dissipation regime of WI is
particularly appealing from both a theoretical and effective field
theory point of view. In particular, the strong dissipation regime of
WI can naturally result in a smaller energy scale inflation with
sub-Planckian field excursions.  These are features that have been
attracting considerable attention, especially more recently, e.g.,
based on the swampland program aiming at finding effective field
theories which can consistently be embedded in quantum gravity theories 
and the role played by WI in achieving this
goal~\cite{Das:2020xmh,Brandenberger:2020oav,Goswami:2019ehb,Berera:2019zdd,Kamali:2019xnt,Das:2019hto,Das:2018rpg,Motaharfar:2018zyb}. This
issue was also partially addressed recently by building two distinct
explicit WI models,  namely the ``variant of warm little inflaton"
(VWLI)~\cite{Bastero-Gil:2019gao} and the ``minimal warm inflation"
(MWI)~\cite{Berghaus:2019whh,Laine:2021ego},  utilizing different
field contents in each construction. In the VWLI, the strong
dissipative regime is obtained by interpolating between decreasing and
growing modes, while in the MWI it is achievable only for some
particular form of potentials, while growing/decreasing modes for the
power spectrum can still exist in both cases.

As far as well-motivated theoretical field theory constructions for
inflation are concerned, there has been considerable progress towards
this goal in recent years, with the construction of several concrete
string inflation models for instance (for reviews, see, e.g.,
Refs.~\cite{Kallosh:2007ig,Baumann:2009ni} and references therein),
with particular attention given to  brane inflation
scenarios\footnote{There are two realizations of brane inflation,
  namely ultraviolet (UV) and infrared (IR) models, depending on
  whether the brane is moving towards or away from the tip of the
  throat. In this paper, we consider an UV model (see
  Refs.~\cite{Chen:2005ad,Chen:2004gc} for IR models).}.  {\it
  Dirac-Born-Infeld (DBI) inflation} is one such string theoretic
motivated model in which the inflaton is interpreted as a modulus
parameter of a D-brane propagating in a warped throat region of an
approximate Calabi-Yau flux
compactification~\cite{Silverstein:2003hf,Alishahiha:2004eh}. Hence,
the effective action for DBI inflation contains a special form of the
DBI kinetic term. The DBI kinetic term introduces some novel speed
limits on the inflaton field velocity and helps in keeping it near the
top of potential, even when it is too steep, thus resulting in the
slow-roll phase through a low sound speed, instead of from dynamical
friction due to expansion. A low sound speed, smaller than unity,
allows the inflaton field to have a sub-Planckian evolution, even for
steep potentials, thus circumventing the
$\eta$-problem~\cite{Easson:2009kk}. Moreover, fluctuations also
propagate with a low sound speed parameter, resulting in both a
smaller tensor-to-scalar ratio and a significant non-Gaussianity,
which can potentially be distinguishable from other
scenarios~\cite{Silverstein:2003hf,Alishahiha:2004eh}.

Although DBI inflation models were greatly successful, it was realized
that the microscopic bound on the maximal field variation due to
compactification is able to put strong constraints on the throat
volume~\cite{Baumann:2006cd} and bulk volume~\cite{Chen:2006hs}. These
issues, together with the Lyth bound, lead to a model independent
upper bound on the tensor-to-scalar ratio which turns out to be
inconsistent with the stringent observational lower bound on
gravitational waves that are produced by the inflationary
dynamics~\cite{Lidsey:2007gq}. Moreover, a viable reheating process,
which typically involves brane/antibrane
annihilation (see, e.g., Refs.~\cite{Barnaby:2004gg,Frey:2005jk,Chialva:2005zy,Firouzjahi:2005qs} for 
earlier studies on reheating in brane inflation), is highly constrained due to
overproduction of long-lived Kaluza-Klein (KK)
modes~\cite{Kofman:2005yz} (note, however, that subsequently Refs.~\cite{Chen:2006ni,Berndsen:2008my,Dufaux:2008br,Frey:2009qb}
argued that warped KK modes can instead serve as dark
matter candidates). 

Reference~\cite{Cai:2010wt} showed that the cold DBI inflation
model can be reconciled with the observations, removing the
aforementioned inconsistencies, provided that the strong dissipation
regime of Dirac-Born-Infeld warm inflation (DBIWI) is achieved. This
is also because the WI scenario violates the Lyth bound. However, the
work done in Ref.~\cite{Cai:2010wt} made use of a  phenomenological
dissipation coefficient, assuming it to be independent of the
temperature\footnote{The stringent observational lower bound on tensor-to-scalar
  ratio found in Ref.~\cite{Lidsey:2007gq} can be relaxed in other DBI setups without dissipation, for
  instance, in multi-brane~\cite{Huston:2008ku} and
  multi-field~\cite{Langlois:2009ej} models.}. However, almost all
explicit particle physics realizations of WI result in an explicit
temperature dependent dissipation
coefficient~\cite{Berera:2008ar,BasteroGil:2010pb,BasteroGil:2012cm,Bastero-Gil:2016qru,Bastero-Gil:2019gao,Berghaus:2019whh}. Besides
this, pushing WI into the strong dissipation regime is challenging
due, again, to the aforementioned problem related to the appearance of
a growing/decreasing function in the power spectrum, which tends to
lead to results inconsistent with the observations, e.g., for the
spectral tilt~\cite{Benetti:2016jhf}. Although there have been some
previous papers considering DBIWI
realizations~\cite{Cai:2010wt,Zhang:2014dja,Rasouli:2018kvy}, none of
them made an investigation of the role of the sound speed on the
backreaction of the radiation perturbations on the inflaton
ones. Thus, none of the earlier references on DBIWI have studied how
the growing/decreasing function in WI gets affected by the sound
speed, which requires a detailed study of the perturbations in DBIWI.
Taken all together, the purpose of this paper is to cover this issue
and to properly understand the primordial perturbations in DBIWI and the
consequences that it brings to model realizations, as far as the
observations constraints are taken into account. Our results show that
DBIWI is able to go into the strong dissipation regime for
well-motivated field theory realizations of WI that result in explicit
temperature dependent dissipation coefficients. Our results also shed
new light on the potential use of DBI type of models in the context of
WI. 

The outline of the remainder of this paper is as follows. In
Sec.~\ref{sec2}, we present the dynamics of DBIWI realization and
investigate the behavior of the dynamical parameters in the
model. Then, in Sec.~\ref{sec3}, we study the perturbation
equations for the DBIWI  and present the backreaction of the thermal
radiation bath on the inflaton perturbations and scalar of curvature
power spectrum. In Sec.~\ref{sec4} we discuss the results obtained
for the DBIWI and demonstrate the effect of a low sound speed on the
spectrum of DBIWI. {}Finally, in Sec.~\ref{conclusions}, we give a
summary of the results and conclude by discussing the implications of
the DBIWI realization from a model building perspective.  Throughout
this paper, we work with the natural units, in which Planck's
constant, the speed of light and Boltzmann's constant are set to
$1$ and we also work with the reduced Planck mass,
$M_{\rm Pl} \equiv 1/\sqrt{8 \pi G} \simeq 2.4 \times 10^{18}$GeV.

\section{DBIWI background dynamics}
\label{sec2}

The dynamical realization of WI is different from the CI one due to
the presence of radiation and energy exchange between the inflaton
field and the radiation energy density. Hence, the total energy
density of the Universe in WI contains both the inflaton field and a
primordial radiation energy density, i.e., $\rho = \rho_{\phi} +
\rho_{r}$, where $\rho_{\phi}$ and $\rho_{r}$ are the inflaton field
and the radiation energy densities, respectively. Even when $\rho_{r}$
is subdominated at the beginning and throughout inflation,  i.e.,
$\rho_{r}\ll \rho_{\phi}$,  the underlying dissipation effects
generating this radiation energy density are still able to modify the
inflationary dynamics and the perturbations in a nontrivial way in WI.
The inflaton field and radiation energy density form a coupled system
in the WI dynamics due to dissipation of energy out of the inflaton
system and into radiation. In the spatially flat
{}Friedmann-Lema\^{\i}tre-Robertson-Walker (FLRW) metric, the background
evolution equations are, respectively, given by
\begin{align}
&\dot \rho_{\phi} + 3 H(\rho_{\phi}+p_{\phi}) = -\Upsilon
  \left(\rho_{\phi}+p_{\phi}\right)\label{inflaton},\\ &\dot \rho_{r}
  + 3 H (\rho_{r} + p_{r}) = \Upsilon
  \left(\rho_{\phi}+p_{\phi}\right)\label{radiation},
\end{align}
where $p_{\phi}$ and $p_{r}$ are the inflaton and radiation pressures,
respectively, and $\Upsilon$ is the dissipation coefficient, which in
general can be a function of both the inflaton field $\phi$ and
temperature $T$ of the produced radiation bath. 

In the DBIWI realization, the inflaton is a  modulus  parameter  of  a
D-brane  moving  in a  warped  throat  region which dissipates its
vacuum energy into radiation through Eqs.~(\ref{inflaton}) and
(\ref{radiation}), while its evolution is effectively governed by the
following Lagrangian density,
\begin{align}
\mathcal{L}_{DBI} =  f^{-1}(\phi) \left[1- \sqrt{1-2f(\phi)X}\right]-
V(\phi),
\end{align}
where $X = - \frac{1}{2} \partial^{\mu}\phi \partial_{\mu}\phi$, with
$V(\phi)$ being the potential function for the inflaton and $f(\phi)$
is the  redefined warp  factor.  The form for the warp factor function
can  be phenomenologicaly  deformed  depending on the desired model
construction. {}For  the  well-studied  anti-de Sitter
throat~\cite{Silverstein:2003hf,Alishahiha:2004eh} the warp factor
function is given by $f(\phi) = {f_{0}}/{\phi^{4}}$ with $f_{0}$ being
a  positive  constant. {}For definiteness, this is the form for the
warp factor function that we will be assuming in this work. 

Varying the action with respect to the metric, the energy density
$\rho_\phi$ and the pressure $p_\phi$ of the DBI field are given,
respectively, by
\begin{align}\label{density}
& \rho_{\phi}=  \frac{\dot\phi^{2}}{c_{s}(1+c_{s})} + V(\phi), \\ &
  p_{\phi} = \frac{\dot \phi^{2}}{1+c_{s}} - V(\phi),
\end{align}
where $c_{s} \equiv \sqrt{1-2Xf(\phi)}= \gamma^{-1}$ is the sound
speed, which is smaller than unity as a consequence of the nontrivial
kinetic structure of the DBI Lagrangian, with $\gamma$ similarly being
the Lorentz factor in five dimensions. Inserting Eq.~(\ref{density})
in Eqs.~(\ref{inflaton}) and (\ref{radiation}), the dynamical equation
for the inflaton field in the DBIWI is found to be given by 
\begin{align}
& \ddot \phi + 3c_{s}^{2}H(1+Q)\dot\phi + c_{s}^{3}V^{\prime}+
  \frac{f^{\prime}}{2f^{2}}\left(1-3c_{s}^{2}+ 2c_{s}^{3}\right) = 0,
\label{phieom}
\end{align}
while for the radiation energy density is
\begin{align}
 &\dot \rho_{r} + 4H\rho_{r} = c_{s}^{-1}\Upsilon {\dot\phi^{2}},
\label{radeom}
\end{align}
where $Q = \Upsilon/(3H)$ gives a measure for the strength of the
dissipative processes in WI. Equation~(\ref{phieom}) for the inflaton
background evolution can also be obtained from the following covariant
equation (when taking $\phi$ as an homogeneous field), 
\begin{eqnarray}\label{covariant}
&& c_{s}^{-1}\Box \phi - c_{s}^{-3} 
f(\phi)(\nabla_{\mu}\nabla_{\nu}\phi)(\nabla^{\mu}\phi\nabla^{\nu}\phi)
- V^{\prime} 
\nonumber \\ 
&& -
\frac{f^{\prime}(\phi)}{2f^{2}(\phi)}\left(c_{s}^{-3}-3c_{s}^{-1}+2\right)
=-c_{s}^{-1}\Upsilon u^{\mu}\partial_{\mu} \phi.
\end{eqnarray}

In the slow-roll regime, $\ddot \phi \ll H\dot\phi$ and
$\dot\rho_{r}\ll 4H\rho_{r}$, i.e., the inflaton is slowly varying and
radiation is produced in a quasistationary way.  Both slow-roll
conditions hold as long as
\begin{equation}
\epsilon_{s} \equiv \frac{d \ln c_{s}}{dN} = \frac{\dot c_{s}}{Hc_{s}}
\ll 1,
\end{equation} 
with $\epsilon_{s}$ quantifying the variation of the sound speed,
along with the usual condition that the standard slow-roll parameters
in WI are small. Then, under the slow-roll approximation, the
background equations reduce to 
\begin{align}\label{dotphi}
\dot\phi \simeq -\frac{c_{s}V^{\prime}}{3H(1+Q)} , \ \   \rho_{r}
\simeq \frac{3Q\dot\phi^{2}}{4c_{s}} , \ \   3H^{2}  \simeq
\frac{V(\phi)}{M_{\rm Pl}}.
\end{align}

To study the model quantitatively, we need to fix the functionality of
the potential and of the dissipation coefficient. In this paper, we
consider a monomial power-law potential for the inflaton,
\begin{equation}
V(\phi) = \frac{V_0}{2n} \left(\frac{\phi}{M_{\rm Pl}}\right)^{2n},
\label{Vphi}
\end{equation}
where $V_{0}$ is the amplitude of the potential and $n$ is a real
number. Moreover, the dissipation coefficient can be very well
parameterized by\footnote{See also
  Refs.~\cite{Zhang:2009ge,Visinelli:2016rhn} for some earlier studies
  also considering this functional form for the dissipation
  coefficient in WI.}
\begin{equation}
\Upsilon(\phi,T)=C_\Upsilon\,T^c \phi^p M^{1-p-c},
\label{Upsilon}
\end{equation}
where $C_\Upsilon$ is a constant, $T$ is the temperature, $M$ is some appropriate mass
scale and both quantities can be associated with the specifics of the
microscopic model parameters leading to Eq.~(\ref{Upsilon}) (see,
e.g.,
Refs.~\cite{Gleiser:1993ea,Berera:1998gx,Berera:2008ar,BasteroGil:2012cm,Bastero-Gil:2016qru,Laine:2021ego}
for some explicit examples of quantum field theory model derivations
for dissipation coefficients used in WI).

The scalar of the curvature power spectrum in
WI~\cite{Graham:2009bf,BasteroGil:2011xd,Bastero-Gil:2014jsa,Ramos:2013nsa}
has strong dependence on the dissipation ratio $Q$ and on the
temperature over the Hubble parameter ratio,
$T/H$. One also notes that strictly speaking
  $T/H$ can also be considered as a function of $Q$, once we use the
  CMB amplitude value to constrain the scalar power spectrum. In this
  case, $Q$ can be considered the relevant quantity parametrizing the
  WI dynamics.  It is then useful to look at the dynamical
behavior of these quantities during inflation. In particular, the
analysis of their behavior is important to determine regimes where the
spectral tilt of the scalar of curvature power spectrum increases
(leading potentially to a blue-tilted spectrum) or decreases (and that
can lead to a red-tilted spectrum).  Such an analysis is particularly
relevant in modeling WI models and constraining them with the
observations~\cite{Benetti:2016jhf}. {}Furthermore, since $Q$ appears
explicitly in the slow-roll parameters of WI, the behavior of $Q$
during WI is important for determining regimes where the dynamics are
consistent and can have a graceful exit~\cite{Das:2020lut}. In the
case of DBIWI, the slow-roll coefficients also depend on the
sound speed $c_s$, hence, for completeness, we also study its behavior
here. During slow-roll we find that the dissipation ratio $Q$, the
temperature over Hubble parameter $T/H$, and the sound speed $c_s$ in
DBIWI evolve with the number of e-folds $N$, respectively, as
\begin{widetext}
\begin{eqnarray}
\frac{d \ln Q}{dN} &=& \frac{c_s \left\{\left[4 + c(1+c_s^2)\right]
  \epsilon_V - c (1 + c_s^2) \eta_V -  (2 c (1 - c_s^2)+ 4 p) \kappa_V
  \right\}} {4 - c + 4 Q + c c_s^2 Q},
\label{dQdN}
\\ \frac{d \ln (T/H)}{dN} &=& \frac{c_s \left\{\left(6 + c_s^2 + 5 Q +
  c (-1 + c_s^2 Q)\right)  \epsilon_V - (1 + c_s^2) (1 + Q) \eta_V +
  \left[p(c_{s}^{2}Q-1) - 2(1+Q) (1-c_{s}^{2}) \right]  \kappa_V
  \right\}}{(1 + Q) (4 - c + 4 Q + c c_s^2 Q)}, \nonumber \\
\label{dTHdN}
\\ \frac{d \ln c_s}{dN} &=& \frac{ c_s (1 - c_s^2) \left\{(-4 + c + c
  Q) \epsilon_V - (-4 + c)  (1 + Q) \eta_V - 2 \left[4 - c + (4 + c +
    2 p) Q\right] \kappa_V\right\} } {(1 + Q) (4 - c + 4 Q + c c_s^2
  Q)},
\label{dcsdN}
\end{eqnarray}
\end{widetext}
where $\epsilon_V = M_{\rm Pl}^2 (V^{\prime}/V)^{2}/2$ and $\eta_V =
M_{\rm Pl}^2 V^{\prime\prime}/V$ are the usual slow-roll inflaton
potential parameters and $\kappa_{V} = M_{\rm
  Pl}^2 V^{\prime}/(\phi \,V)$.

We note from Eqs.~(\ref{dQdN})-(\ref{dcsdN}) that the denominator of
those expressions is always positive.  This is because the power $c$
in the temperature dependence of the dissipation coefficient satisfies
$-4 < c < 4$ (see, e.g.,
Refs.~\cite{Moss:2008yb,delCampo:2010by,BasteroGil:2012zr}).  On the
other hand, the sign of the numerator in the above expressions will
depend on both the form of the  dissipation coefficient and on the
inflaton potential exponent $n$.  As examples of representative cases
of WI models, we can consider two cases of dissipation coefficient
that are well motivated microscopically: (a) $c=1,\,p=0$, i.e., a
linear in the temperature dependence for the dissipation coefficient,
$\Upsilon \propto T$, that was first derived in
Ref.~\cite{Bastero-Gil:2016qru} in the case of the WLI; and (b)
$c=3,\,p=0$, i.e., a cubic power in the temperature dependence for the
dissipation coefficient, $\Upsilon \propto T^3$, and derived recently
in Refs.~\cite{Berghaus:2019whh,Laine:2021ego} in the case of the MWI.
{}From the Eqs.~(\ref{dQdN})-(\ref{dcsdN}) we find that in case (a)
both $Q$ and $T/H$ are growing  functions with the number of e-folds
for $n>1/3$ for all $c_s$, while $c_s$ initially decreases for $n<3$
when $Q \ll 1$, and for $Q> 1$ it will increase for $n>13/7$ and
decreases otherwise.  In case (b) $Q$ and $c_s$  are decreasing
functions with the number of e-folds whenever $n<3$ and $c_s^2<1/3$ or
$n>3$ and $c_s^2> 1/3$ and increasing functions with $N$ otherwise,
while $T/H$ is in general a growing function with $N$. Both cases (a)
and (b) then reinforce the WI condition $T/H >1$ throughout the
inflationary dynamics.

\section{Perturbations for DBIWI}
\label{sec3}

Let us describe the scalar perturbation equations in DBIWI.  We start
with the fully perturbed FLRW metric,
\begin{eqnarray}
ds^{2}& =& -(1+2\alpha)dt^{2}-2a\partial_{i}\beta dx^{i}dt  \nonumber
\\  &+& a^{2}[\delta_{ij}(1+2\varphi) +
  2\partial_{i}\partial_{j}\gamma]dx^{i}dx^{j},
\end{eqnarray} 
where $\alpha$, $\beta$, $\gamma$ and $\varphi$ are the
spacetime-dependent metric perturbation variables.  In WI the
evolution of field fluctuations $\delta \phi$, which is effectively
described by a stochastic evolution, is determined by a Langevin-like
equation~\cite{Graham:2009bf,Bastero-Gil:2014jsa}.   In the case of
DBIWI, perturbing the covariant equation~(\ref{covariant}) around the
background, i.e., $\Phi({\bf x},t) = \phi(t) + \delta\phi({\bf x},
t)$, the corresponding Langevin equation for $\delta \phi$ reads
\begin{widetext}
\begin{align}
\nonumber \delta\ddot\phi &+ 3 H \ \left(1 +Q-
\epsilon_{s}\right)\delta\dot\phi +  H^{2} \left\{ z^{2}+ \frac{3\dot
  f}{2Hf} \left[ (1-c_{s}^{2}) (1+Q) - \epsilon_{s}\right]-
\frac{1-3c_{s}^{2}+2c_{s}^{3}}{H^{2}}\left(\frac{{f^{\prime}}^{2}}{f^{3}}
-  \frac{f^{\prime\prime}}{2f^{2}} \right) +
\frac{c_{s}^{3}V^{\prime\prime}}{H^{2}} \right\} \delta\phi  \\ & +
c_{s}^{2}\delta \Upsilon \dot \phi =c_{s}^{3}\left(\xi_q
+\xi_\Upsilon\right) + \left[(3c_{s}^{-2}-1)\ddot \phi +
  3H(1+Q)\dot\phi\right] \alpha +  (\dot\alpha+  c_{s}^{2} \kappa)\dot
\phi - \frac{3f^{\prime}}{2f^{2}}(1-c_{s}^{2})(1-c_{s}^{-2})\alpha, 
\label{deltaddotphi}
\end{align}
\end{widetext}
which is also supplemented by the first-order perturbation equations
for the radiation energy density, $\delta \rho_r$, and for the
radiation momentum perturbation, $\Psi_r$, given, respectively, by
\begin{align}
\delta \dot \rho_r & + 4 H \delta \rho_r = \frac{k^2}{a^2} \Psi_r +
\dot \rho_{r} \alpha+ \frac{4}{3}\rho_{r}\kappa + \delta Q_r, 
\label{deltarhor}
\end{align}
and 
\begin{align}
\dot \Psi_r & + 3 H\Psi_r  = -\frac{\delta \rho_r}{3}-
\frac{4}{3}\rho_{r}\alpha + J_r.
\label{deltaPsir}
\end{align}

In the Eqs.~(\ref{deltaddotphi}), (\ref{deltarhor}) and
(\ref{deltaPsir}), we have that
\begin{eqnarray}
\delta \Upsilon &=& 3HQ \left(\frac{c\delta{\rho_{r}}}{4\rho_{r}} +
\frac{p\delta\phi}{\phi}\right),
\label{deltaUpislon}
\\ \nonumber \\ \delta {Q}_{r} &=& c_{s}^{-1}\delta
\Upsilon\dot\phi^{2} + 3 H^{2}Q c_{s}^{-3}(1-c_{s}^{2})  \frac{\dot
  f}{2Hf} \dot\phi\delta\phi \nonumber \\ &+&
3c_{s}^{-3}(1+c_{s}^{2})HQ \dot\phi\delta\dot\phi \nonumber \\ &-&
3c_{s}^{-3}(1+c_{s}^{2})HQ \dot\phi^{2}\alpha,
\label{deltaQr}
\\ \nonumber \\ J_{r} &=& - c_{s}^{-1}\Upsilon \dot \phi \delta\phi,
\label{Jr}
\end{eqnarray}
where $z= c_{s}k/(aH)$, $\kappa =
3(H\alpha-\dot\varphi)+k^{2}\chi/a^{2}$ and $\chi =
a(\beta+a\dot\gamma)$.   Moreover, in Eq.~(\ref{deltaddotphi})
$\xi_{q,\Upsilon}\equiv \xi_{q,\Upsilon}({\bf k},t)$  are stochastic
Gaussian sources related to quantum and thermal fluctuations with
appropriate amplitudes chosen such as to match the analytical
derivation for the scalar of curvature power spectrum in
WI~\cite{Ramos:2013nsa}. This leads in particular that
$\xi_{\Upsilon}$ and $\xi_{q}$ both having zero mean and satisfying
two-point correlation functions given, respectively, by
\begin{eqnarray}
\!\!\!\!\!\!\!\!\!\!\langle \xi_\Upsilon({\bf k},t)\xi_\Upsilon({\bf
  k}',t')\rangle &=& \frac{2 \Upsilon T}{a^3} \delta(t-t') (2 \pi)^3
\delta({\bf k} + {\bf k}'), 
\label{xiUpsilon}
\\  \!\!\!\!\!\!\!\!\!\!\langle \xi_q({\bf k},t)\xi_q({\bf
  k}',t')\rangle &=& \frac{H^2(9+12\pi Q)^{1/2}(1+2n_*)}{\pi a^3}
\nonumber \\ &\times& \delta(t-t') (2 \pi)^3 \delta({\bf k} + {\bf
  k}'),
\label{xiq}
\end{eqnarray}
where $n_*$ represents the statistical distribution state for the
inflaton quanta at the Hubble radius~\cite{Ramos:2013nsa} (see also
the Appendix B of Ref.~\cite{Das:2020xmh} for details). 

The set of perturbation equations (\ref{deltaddotphi}),
(\ref{deltarhor}) and (\ref{deltaPsir}) are gauge-ready
equations. They can be used with any appropriate gauge choice or also
be  worked out in terms of gauge-invariant
quantities~\cite{Kodama:1985bj,Hwang:1991aj}.  {}For instance, they
can be taken in the Newtonian-slicing (or zero-shear) gauge $\chi=0$,
with the relevant metric equations becoming
\begin{eqnarray}
&&\kappa= \frac{3}{2 M_{\rm Pl}^2} ( c_s^{-1} \dot{\phi} \delta \phi -
  \Psi_r )\,,
\label{kappachi=0}\\
&& \alpha = -\varphi \,,
\label{alphachi=0}\\
&& \dot{\varphi} = - H \varphi -\frac{1}{3} \kappa\,.
\label{dotvarphichi=0}
\end{eqnarray}

It can also be easily checked that Eq.~(\ref{deltaddotphi}) reduces to
the standard Langevin equation in
WI~\cite{Graham:2009bf,Bastero-Gil:2014jsa} for $c_{s}=1$, i.e.,
$\epsilon_{s}=0$.  As it is clear from Eqs.~(\ref{deltaddotphi}) and
(\ref{deltarhor}), the term $\delta \Upsilon$ is responsible for
coupling the inflaton perturbations with those of the radiation
whenever $c\neq 0$ and which is known to lead to a growing mode for
the resulting power spectrum~\cite{Graham:2009bf}. 

\subsection{The scalar of curvature power spectrum}

{}Given the perturbation equations,
Eqs.~(\ref{deltaddotphi})-(\ref{deltaPsir}), the scalar power spectrum
is determined from the comoving curvature perturbation  ${\cal R}$, 
\begin{equation}
\Delta_{\cal R}(k)= \frac{k^3}{2 \pi^2} \langle |{\cal R}|^2
\rangle\,, \label{PR}
\end{equation}
where ``$\langle \cdots \rangle$'' means averaging over different
realizations of the noise terms in Eq.~(\ref{deltaddotphi}) (see, for
instance
Refs.~\cite{Graham:2009bf,BasteroGil:2011xd,Bastero-Gil:2014jsa} for
details of the numerical procedure). Given an appropriate gauge,
${\cal R}$ is composed of contributions from the inflaton momentum
perturbations $\Psi_\phi = -c_{s}^{-1}\dot \phi \delta \phi$ and from
the radiation  momentum perturbations $\Psi_r$,
\begin{eqnarray}
&&{\cal R}= \sum_{i=\phi,r} \frac{\rho_i + p_i}{\rho+p}{\cal R}_i\;,
\label{R}\\
&&{\cal R}_i =  -\varphi- \frac{H}{\rho_i + p_i} \Psi_i\,,
\label{Ri}
\end{eqnarray}
with $p=p_\phi + p_r$,  $\rho_\phi+p_\phi =c_{s}^{-1}\dot\phi^2$ and
$\rho_r+p_r \simeq 4 \rho_r/3 = c_{s}^{-1}Q\dot\phi^{2}$.

An explicit analytic expression for the scalar of curvature power
spectrum can be obtained when neglecting the coupling between inflaton
and radiation perturbations (i.e., taking $c=0$ in
Eq.~(\ref{deltaUpislon})) and it was determined in
Ref.~\cite{Ramos:2013nsa}.  The result in this case is
well approximated by~\cite{Ramos:2013nsa}
\begin{equation} 
\Delta_{{\cal R}} \simeq \left(\frac{ H_{*}^2}{2
  \pi\dot{\phi}_*}\right)^2  \left(1+2{n_{BE}} + \frac{2\sqrt{3}\pi
  Q_*}{\sqrt{3+4\pi Q_*}}{T_*\over H_*}\right) ,
  \label{Pphi2}
\end{equation}
where $n_{BE}$ is the Bose-Einstein distribution and the effect of
sound speed parameter is encoded in the inflaton field velocity
through Eq.~(\ref{dotphi}).  One should note that in DBIWI,
fluctuations are frozen at the sound horizon, where
${c_{s}}_{*}k_{*}=a_{*}H_{*}$, rather than at the Hubble radius, where
$k_*=a_*H_*$.  In general we can also replace $n_{BE}$ in
Eq.~(\ref{Pphi2}) by $n_*$, representing the statistical distribution
state of the inflaton at the sound horizon, which might not be
necessarily that of thermal equilibrium.  The form given by
Eq.~(\ref{Pphi2}) is typically the result used in most of the recent
literature in WI when the growing function is absent.  

We want to first understand the effect of the coupling
between inflaton and radiation perturbations and how a sound speed
$c_s <1$ might change the results for the power spectrum in DBIWI.  To
possibly do something analytical and to obtain a feeling for these
effects, we can start by considering some simplifying assumptions.
Since we are interested in the strong dissipation regime $Q>1$, we can
start by neglecting the contribution of quantum  fluctuations to the
power spectrum. {}For $Q > 1$, the contribution from the thermal
stochastic fluctuations $\xi_\Upsilon$ in Eq.~(\ref{deltaddotphi})
dominates over that from the quantum fluctuations $\xi_q$.  Moreover,
we can also analytically solve Eq.~(\ref{deltaddotphi}) explicitly by
neglecting the slow-roll order corrections (and likewise the metric
contributions, which also give slow-order corrections only).  Under
these simplifying assumptions, the authors in
Ref.~\cite{Graham:2009bf} have explicitly shown that the power
spectrum grows with $Q_*$ like $\Delta_{\mathcal{R}} \propto
\Delta_{{\cal R},c=0} Q_*^{3c}$, with 
\begin{equation}
\Delta_{{\cal R},c=0} \simeq
\frac{\sqrt{3}}{4\pi^{\frac32}}\frac{H^{3}_{*}T_{*}}{\dot\phi^{2}_{*}}
Q_{*}^{\frac{1}{2}},
\label{Pphi2Qlarge}
\end{equation}
when taking $Q_* \gg 1$ in Eq.~(\ref{Pphi2}).  By utilizing the same
techniques developed in Ref.~\cite{Graham:2009bf}, but now adapted to
the perturbation equations in DBIWI, we arrive at the
result, when $Q_* \gg 1$, that   $\Delta_{\mathcal{R}}$ is now given
by
\begin{align}
\Delta_{\mathcal{R}} \simeq \Delta_{{\cal R},c=0}
\left(\frac{Q_{*}}{Q_{c}}\right)^{3cc_{s*}^{2}},
\label{spectrumDBIWI}
\end{align}
where $Q_{c}$ is a function of both $c$ and $c_{s}$. The result given
by Eq.~(\ref{spectrumDBIWI}) explicitly shows that $c_s<1$ can
compensate for the result of a growing scalar spectrum amplitude the
larger is $Q$ and whenever $c >0$ (or, likewise, a decreasing
amplitude when $c<0$). The approximations assumed in
Ref.~\cite{Graham:2009bf} do allow us to solve for the inflaton
perturbations and, hence, they help to find an analytical expression
for $\Delta_{\mathcal{R}}$, it has been shown in
Ref.~\cite{BasteroGil:2011xd} that these approximations tend to
overestimate the effect of the coupling between the inflaton and
radiation perturbations. By accounting for the full expressions, i.e.,
not dropping all slow-roll order terms in the equations, the
dependence of  $\Delta_{\mathcal{R}}$ on $Q_*$ tends to be more
suppressed, with a smaller power in $Q_*$. Hence, the results from the
(numerical) solution for the full perturbation equations shows that
$\Delta_{\mathcal{R}} \propto Q_*^\beta$, with $\beta < 3 c$, when
$c_s=1$. This result is also observed in our analysis shown in the
next section, showing that $\beta < 3 c c_{s*}^2$ also holds here in
the DBIWI. However, the effect is still large enough to make the
spectrum to depart considerably from the result given by
Eq.~(\ref{Pphi2}).     By explicitly numerically solving the set of
perturbations equations, as in Ref.~\cite{BasteroGil:2011xd}, the
result of the coupling between inflaton and radiation perturbations
can be expressed as an overall correction to Eq.~(\ref{Pphi2}), which
modifies it to
\begin{equation} 
\Delta_{{\cal R}} =  \left(\frac{ H_{*}^2}{2 \pi\dot{\phi}_*}\right)^2
\left(1+2n_{ BE} + \frac{2\sqrt{3}\pi Q_*}{\sqrt{3+4\pi Q_*}}{T_*\over
  H_*}\right) G(Q_*),
  \label{Pk}
\end{equation}
where $G(Q_*)$ accounts for the effect of the coupling of the inflaton
and radiation fluctuations.  Some explicit forms for $G(Q_*)$ have
been given in the literature, depending on the form of the dissipation
coefficient in
WI~\cite{BasteroGil:2011xd,Bastero-Gil:2014jsa,Benetti:2016jhf,Bastero-Gil:2016qru,Motaharfar:2018zyb,Kamali:2019xnt,Lima:2019yyv,Das:2020xmh}.
In the next section we will give results for $G(Q_*)$ in the DBIWI
case for some of the most representative dissipation coefficient forms
used in recent literature and we will also explicitly see how a
$c_s<1$ effectively suppresses the effects of a growing $Q_*$ in the
amplitude of the scalar power spectrum.  But before going into that
analysis, let us also derive some useful results concerning the
spectral tilt in DBIWI.

\subsection{The spectral tilt $n_s$ in DBIWI}

{}From Eq.~(\ref{Pk}) and also using Eqs. (\ref{dQdN})-(\ref{dcsdN}),
we can explicitly find expressions for the spectral tilt $n_s$,
\begin{equation}
n_s-1 =\frac{d \ln \Delta_{{\cal R}}(k)}{d \ln k} \simeq \frac{d \ln
  \Delta_{{\cal R}}(k)}{d N},
\end{equation} 
which in the weak and strong dissipation regimes of DBIWI are given,
respectively, by
\begin{eqnarray}
\left(n_s-1\right)\Bigr|_{Q_*\ll 1} &=&
\frac{c_{s*}}{4-c}\left\{-\left[10+7 c_{s*}^2-c(3+2c_{s*}^2) \right]
\epsilon_V \right.  \nonumber \\ &+& \left.  \left[ -1 + (7-2c)
  c_{s*}^2 \right] \eta_V \right. \nonumber \\ & + & \left. \left[
  (14-4c)(1-c_{s*}^2) - p \right] \kappa_V \right\},
\label{nsQsmall}
\end{eqnarray}
and
\begin{eqnarray}
\left(n_s-1\right)\Bigr|_{Q_*\gg 1} &&= -\frac{c_{s*}}{2(4+c c_{s*}^2)
  Q_*} \left\{ \left[ 18-c (1-c_{s*}^2)\right] \epsilon_V \right.
\nonumber \\  && \!\!\!\!\!\!\! + \left. \left[2+ c + (-14+5c)
  c_{s*}^2 \right] \eta_V \right.   \nonumber\\  &&  \!\!\!\!\!\!\! +
\left. 2\left[ c (1-c_{s*}^2) + 2 (-7+p) + 7 (2+p) c_{s*}^2 \right]
\kappa_V \right\}   \nonumber \\  && \!\!\!\!\!\!\! +
\frac{c_{s*}}{4+c c_{s*}^2}\left\{\left(4+c+c c_{s*}^2\right)
\epsilon_V - c (1+c_{s*}^2) \eta_V \right.  \nonumber \\  &&
\!\!\!\!\!\!\! -  \left. \left[ 2c(1-c_{s*}^2) + 4 p \right] \kappa_V
\right\} \frac{G'(Q_*)}{G(Q_*)}.
\label{nsQlarge}
\end{eqnarray}
Note that in the case of the primordial inflaton potential of the form
Eq.~(\ref{Vphi}), we have in  Eqs.~(\ref{nsQsmall}) and
(\ref{nsQlarge}) that
\begin{eqnarray}
\epsilon_V &=& \frac{2 n^2 M_{\rm Pl}^2 }{\phi_*^2}, \\ \eta_V &=&
\frac{2n (2n-1)M_{\rm Pl}^2}{\phi_*^2}, \\ \kappa_V &=& \frac{2 n
  M_{\rm Pl}^2}{\phi_*^2}.
\end{eqnarray}
In deriving the Eqs.~(\ref{nsQsmall}) and (\ref{nsQlarge}) for $n_s$
we have also considered that in WI that $T_*/H_* \gg 1$ (in
particular, $T_*/H_* \gg 1$ can be explicitly  verified in the
examples to be considered in the next section). {}Finally, in the
regime $T_*/H_* \ll 1$ and $Q_* \ll1$ we recover the cold inflation
standard result, $n_s-1=-6 \epsilon_V + 2 \eta_V$.  We can note from
Eq.~(\ref{nsQlarge}) that when $c_{s*} \ll 1$, in the absence of the
growing mode, $G'(Q) \to 0$, we have that  $n_s-1 \propto - 1/Q$ and
the spectral tilt will tend to be red-tilted for large $Q$. However,
in the presence of the growing mode, the last term in
Eq.~(\ref{nsQlarge}) dominates at large $Q$ and it will tend to drive
the spectrum to be blue-tilted (for $G'(Q) > 0$).  To better
illustrate the latter case, let us consider a few relevant WI
dissipation terms that have commonly been used in the literature,
namely, (a) the linear in the temperature dissipation
coefficient~\cite{Bastero-Gil:2016qru}, with $c=1,\, p=0$; (b) the
cubic in the temperature  dissipation coefficient, with $c=3,\,p=0$
(e.g., from Ref.~\cite{Berghaus:2019whh}) and (c) the dissipation
coefficient with $c=3,\, p=-2$ (see, e.g.,
Refs.~\cite{BasteroGil:2012cm,Bartrum:2013fia}).  {}For definiteness,
we will also consider the example of a primordial potential with a
quartic inflaton potential ($n=2$), which has been the primordial
potential mostly  considered with these dissipation coefficients in
WI.  We will also assume that the growing mode function is well
described by a polynomial function in the dissipation ratio $Q$, with
the leading term for $Q\gg 1$ of the form $G(Q) \propto Q^\beta$. Let
us see the results for each of these three cases separately.

\subsubsection{Case (a)}

{}For case (a), with $c=1,\, p=0$, we obtain for Eq.~(\ref{nsQlarge})
that
\begin{equation}
\left(n_s-1\right)\Bigr|_{Q_*\gg 1} \simeq \frac{-2 c_{s*}
  \left[(17+c_{s*}^2) - 2\beta (5+c_{s*}^2) \right]} { (4+c_{s*}^2)
  Q_* (\phi_*/M_{\rm Pl})^2  }.
\label{nsc=1p=0}
\end{equation}
Hence, for $c_{s*} \ll 1$ we have that the spectral tilt will turn
blue, i.e., $n_s-1>0$, if $\beta > 1.7$, while for $c_{s*}=1$, the
spectral tilt is blue for $\beta > 1.5$.  We recall that for this
case~\cite{Bastero-Gil:2016qru} we have that $\beta \simeq 2.315$ (for
$c_{s*}=1$), thus, for large $Q$ this dissipation coefficient with a
quartic inflaton potential will always disagree with the
observations. But from Eq.~(\ref{spectrumDBIWI}), we expect that
$\beta$ will decrease proportional  to $c_s^2$. Thus, we need at least
$c_{s*}^2 \lesssim 1.7/2.315 \sim 0.7$ for the model to be able to
sustain  a large dissipation DBIWI regime and to have a red-tilted
scalar of curvature power spectrum, consistent with the
observations. This expectation will be confirmed by our numerical
results shown in Sec.~\ref{sec4}.

\subsubsection{Case (b)}
\label{caseb}

{}For case (b), with $c=3,\, p=0$, we obtain for Eq.~(\ref{nsQlarge})
that
\begin{equation}
\left(n_s-1\right)\Bigr|_{Q_*\gg 1} \simeq \frac{-2 c_{s*}
  \left[(23+31c_{s*}^2) - 2\beta (-1+3c_{s*}^2) \right]} { (4+3
  c_{s*}^2) Q_* (\phi_*/M_{\rm Pl})^2  }.
\label{nsc=3p=0}
\end{equation}
Hence, for $c_{s*} \ll 1$ we have that the spectral tilt, different
from the previous case, will always be red, since in this case $n_s-1
\propto -(23 + 2 \beta)$.  When $c_{s*}=1$, we see from
Eq.~(\ref{nsc=3p=0}) that $n_s-1 \propto -(54 - 4 \beta)$ and the
spectral tilt is, thus blue for $\beta > 13.5$.  But in this
case~\cite{Benetti:2016jhf} we have that $\beta \simeq 4.33$ (for
$c_{s*}=1$), hence, for large $Q$ this dissipation coefficient with a
quartic inflaton potential can quite robustly sustain a red-tilted
spectrum. In fact, the effect of having a $c_{s*} <1$ here can
potentially make the spectrum too red tilted to be consistent with the
observations, since a value of $c_{s*}< 1$ would make the exponent
$\beta$ smaller and lead to a redder spectrum. This is confirmed by
our numerical results shown in Sec.~\ref{sec4}.  Since this type of
dissipation coefficient already leads to quite satisfactory results
when compared to the  observations~\cite{Berghaus:2019whh}, even in
the strong dissipation regime of WI, it does not benefit from a DBIWI
construction, at least in the context of monomial chaotic inflaton
potentials.

\subsubsection{Case (c)}

It is useful to compare case (b) with the results for
case (c), when $c=3,\, p=-2$, in which case the dissipation
coefficient depends on the temperature but also on the inflaton
amplitude. In this case, we obtain for Eq.~(\ref{nsQlarge}) that
\begin{equation}
\left(n_s-1\right)\Bigr|_{Q_*\gg 1} \simeq \frac{-2 c_{s*}
  \left[3(5+c_{s*}^2) - 2\beta (7+3c_{s*}^2) \right]} { (4+3 c_{s*}^2)
  Q_* (\phi_*/M_{\rm Pl})^2  }.
\label{nsc=3p=-2}
\end{equation}
Here, we have that for $c_{s*} \ll 1$ the spectral tilt is such that
it will turn blue when $\beta > 15/14 \simeq 1.07$, while for
$c_{s*}=1$, the spectral tilt is blue already for $\beta > 0.9$.  
In this case we still have $\beta \simeq 4.33$ (for $c_{s*}=1$), thus,
for large $Q$ the spectrum will always be quite blue tilted. This
agrees with the previous results for this type of WI
model~\cite{Bartrum:2013fia,Benetti:2016jhf}, which have shown that
this dissipation coefficient, within the mononial chaotic inflaton
potential models, can only be consistent with the observational
data\footnote{{}From the Planck Collaboration~\cite{Akrami:2018odb},
  the result for the spectral tilt is $n_s=0.9658\pm 0.0040$  (95$\%$
  CL, Planck TT,TE,EE+lowE+lensing+BK15+BAO+running) at pivot scale
  $k_*=0.05\,{\rm Mpc}^{-1}$. } when $Q_* \ll 1$, i.e., in the weak
dissipation regime of WI. In this regime of $Q_* \ll 1$, the growing
mode is not an issue, since $G(Q_*) \to 1$.  As in case (a) with a
linear temperature dissipation coefficient, this model can
benefit from  a DBIWI realization by exploring the effect of a small
$c_s$ value, which will help to suppress the growing mode by
decreasing $\beta$. However, this is expected to only happen for a
smaller value for $c_s$ than the one shown for the case (a), $c_{s*}^2
\lesssim  0.2$.  This case also illustrates well the importance in
deriving the full dependence of the dissipation coefficient in both
the temperature and in the inflaton field amplitude. Despite cases (b)
and (c) have exactly the same temperature dependence, the fact that
the dissipation coefficient in (c) has an explicit dependence on
$\phi$, implies that an agreement with the observational value for
the tilt of the scalar power spectrum can only be achieved in the weak
dissipation regime of WI (for the standard case of $c_{s*}=1$), while
in (b) we can robustly support the strong dissipation regime of WI, even
with the presence of the growing mode ()which in that case is benign).

\subsection{Non-Gaussianities in DBIWI}

Before ending this section. Let us briefly comment on the expected
non-Gaussianities in DBIWI.  In the DBIWI realization,
non-Gaussianities may be generated by both nonequilibrium dissipative
effects and by a low sound speed. It was shown in
Ref.~\cite{Moss:2011qc,Bastero-Gil:2014raa} that non-Gaussianities
produced by temperature dependent dissipation coefficients are larger
than temperature independent dissipation coefficients due to the
backreaction of radiation bath and coupling of the inflaton to the
radiation field perturbations. However, we qualitatively expect that
for low sound speeds, $c_{s*} <1$, a suppressing effect, as seen for
the scalar of curvature power spectrum, will also work in
the case of the nonlinear parameter $f_{NL}$. Moreover, the
non-Gaussianity parameter for WI in the strong dissipation regime is
roughly less than $10$, i.e., $|f_{NL}^{Warm}|\leq 10$, for
temperature dependent dissipation
coefficients~\cite{Moss:2011qc,Bastero-Gil:2014raa}. {}Furthermore,
the non-Gaussianity produced by cold DBI inflation is inversely
proportional to the square of the sound speed, i.e., $f_{NL}\simeq
\frac{35}{108} (c_{s}^{-2}-1)$. Hence, the corresponding
non-Gaussianity parameter in cold DBI inflation is $f_{NL} \simeq 3$
for $c_{s*}=0.3$, which is not very large.  Therefore, depending on
how these two non-Gaussianity sources contribute to the
non-Gaussianity parameter (where they can either counterbalance or
enforce each other), we may obtain large or small non-Gaussianities in
DBIWI~\cite{Cai:2010wt, Zhang:2014kwa}. Nonetheless, a comprehensive
analysis is needed to find how large the non-Gaussianity will be in
the DBIWI realization. 

\section{Results}
\label{sec4}

\begin{center}
\begin{figure}[!htb]
\subfigure[\ Linear in $T$ dissipation
  coefficient]{\includegraphics[width=8.2cm]{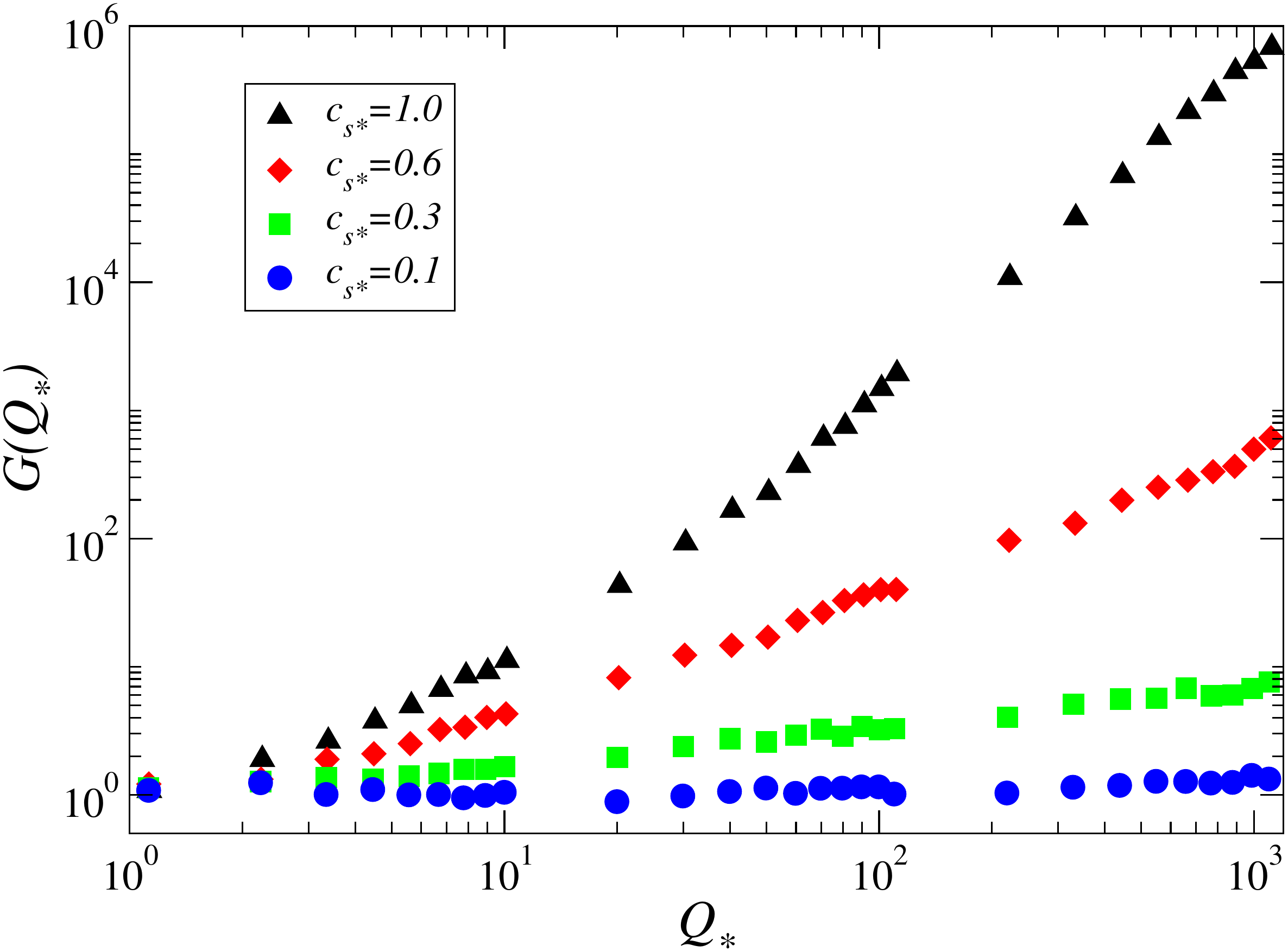}}
\subfigure[\ Cubic in $T$ dissipation
  coefficient]{\includegraphics[width=8.2cm]{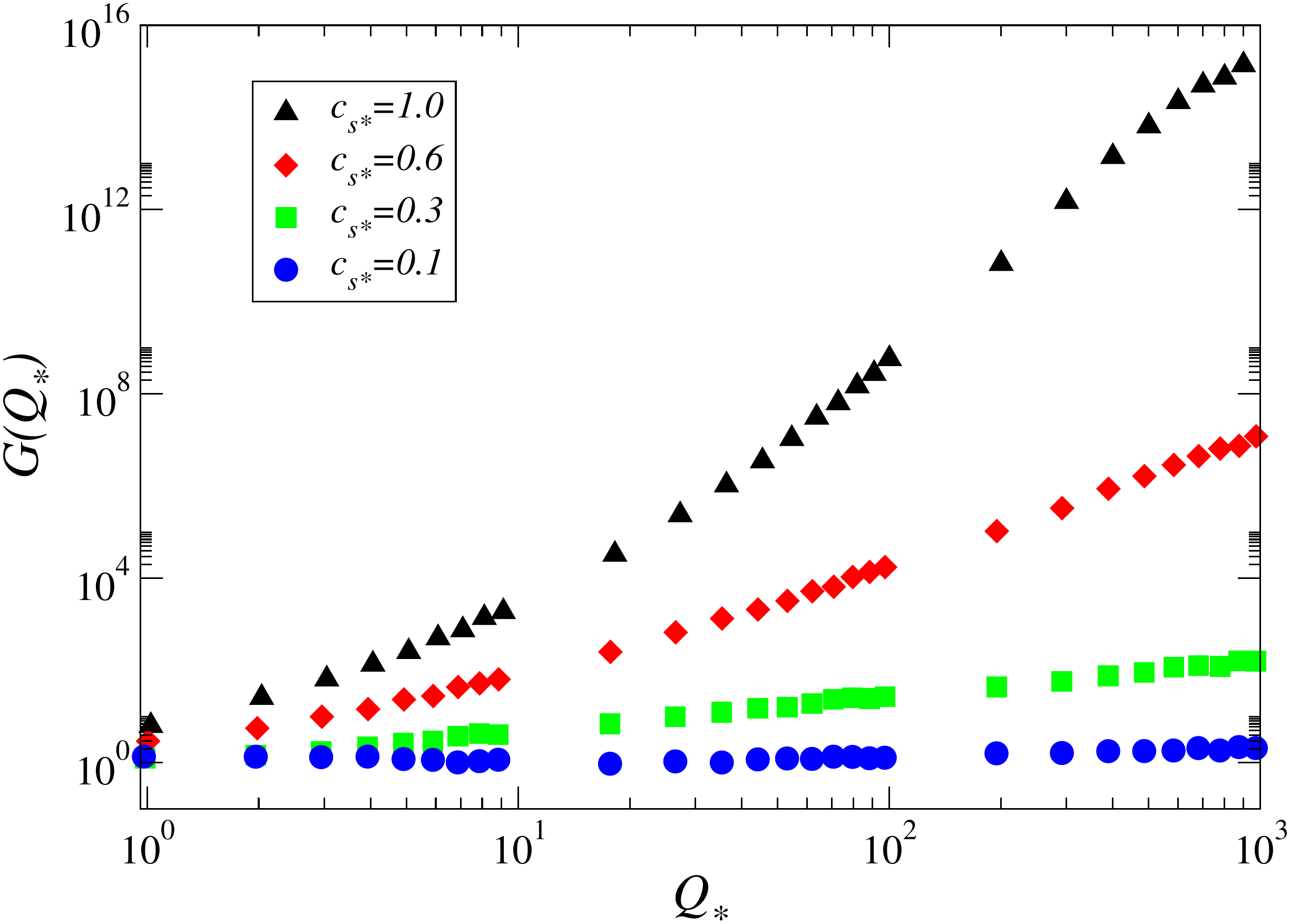}}
\caption{The growing function $G(Q_*)$ as defined by Eq.~(\ref{GQ*}).}
\label{fig1}
\end{figure}
\end{center}

To demonstrate the effect of $c_s$ on the power spectrum in the DBIWI
realization, more precisely on the function $G(Q_{*})$,   we have
considered the quartic  monomial inflaton potential ($n=2$) and
focused our studies for the two representative more recent cases of WI
dissipation coefficients, namely, $\Upsilon = C_{\Upsilon} T$ and
$\Upsilon = C_{\Upsilon} T^3/M^2$.  Results for other cases of
monomial inflaton potential are found to be very similar to the
quartic one, so we refrain from showing those other cases with
expected similar results here. The growing function $G(Q_*)$ is
determined by the numerical evaluation of the perturbation equations
and obtaining the scalar of curvature power spectrum~(\ref{PR}), with
$G(Q_*)$ defined as
\begin{equation}
G(Q_*) = \frac{\Delta_{{\cal R},c\neq 0}}{\Delta_{{\cal R},c= 0}},
\label{GQ*}
\end{equation}
where $\Delta_{{\cal R},c\neq 0}$ is the solution for the scalar of
curvature power spectrum  with the explicit coupling of the inflaton
perturbations with the radiation ones in Eq.~(\ref{deltaddotphi}) and
$\Delta_{{\cal R},c= 0}$ is the solution when this coupling is
explicitly dropped from the equation. We should note that during the
numerical analysis, although we dropped the metric perturbations,
which is justified in an appropriate gauge and in the strong
dissipation regime,  we take into account the effect of first order
slow-roll parameters,  i.e. $\epsilon_{V}$, $\eta_{V}$, $\kappa_{V}$,
etc., in Eq.~(\ref{deltaddotphi})  to obtain the precise behavior
for the growing function.  The results for the linear and cubic
dependencies in temperature for the dissipation coefficient are shown
in {}Fig.~\ref{fig1}.  The results shown in {}Fig.~\ref{fig1} indicate
that for $c_s \lesssim 0.1$ the growing function $G(Q_*)$ can indeed
be taken as $G(Q_*) \approx 1$.

As shown in the previous section, we may obtain a red-tilted spectral
index in the strong dissipation regime even if $c_{s*}$ has
intermediate values, not necessarily for values of $c_{s*} \ll
1$. This is because as $c_{s*} <1$, it can already suffice to
suppress the growing mode enough to allow for a red-tilted spectrum,
even for large $Q$ values. Thus small,  but still reasonable values
for $c_{s*}$ can work to allow the spectrum in DBIWI to be red-tilted
in the strong dissipation regime. Hence, it is useful to obtain the
exact functionality of $G(Q_{*})$ for future analysis. In this regard,
we present the following fitting functions to the curves shown in
{}Fig.~\ref{fig1},
\begin{align}\label{Growingmode}
G(Q_{*}) &= 1+ A_{c} Q^{\alpha}_{*} + B_{c} Q^{\beta}_{*},
\end{align}
where the coefficients $\alpha$, $\beta$, $A_{c}$ and $B_{c}$ are
given in the Table~\ref{coefficients} for the two representative
values of $c$  (i.e., for a linear and a cubic in $T$ dissipation
coefficients, as obtained in the LWI and MWI models, respectively) and
also for three representative  values for $c_{s}$ (at the effective
Hubble radius crossing).

\begin{center}
\begin{table}[!htb]
\caption{The coefficients $\alpha$, $\beta$, $A_{c}$ and $B_{c}$ in growing function $G(Q_{*})$ 
with different values for $c$ and $c_{s}$.}
\begin{tabular}{c|c|c|c|c|c}
\hline\hline
$c$&$c_{s*}$&$\alpha$ &$\beta$ &$A_{c}$&$B_{c}$\\
\hline
 \ \ \ \ & \ \ 1 \ \  &  \ \ 1.364 \ \  &  \ \ 2.315 \ \ & \ \ 0.335 \ \ & \ \ \ 0.0185 \ \ \\
1 & 0.6 & 0.694 & 1.114 & 0.311 & 0.187\\
& 0.3 & 0.395 & 0.448 & 0.205 & 0.127\\
\hline
& 1& 1.946 & 4.330& 4.981& 0.127\\
3 &  0.6 & 1.975 & 2.684 & 0.475 & 0.083 \\   
& 0.3  & 0.815 & 0.939 & 0.478 & 0.368 \\     
\hline\hline 
\end{tabular}
\label{coefficients}
\end{table}
\end{center}

When these results are applied to the cubic in the temperature form
for the dissipation coefficient ($c=3,\,p=0$) and as already
anticipated from the discussion given for case (b) in
Sec.~\ref{caseb}, for values of $c_s < 1$,  we get that $n_s \simeq
0.95$ and even smaller values\footnote{More specifically, for $c_{s*}=1$,
we get for the model with $n=2,\, c=3,\, p=0$ that $n_s \simeq 0.97$
within the whole range of $Q_* \sim 1 - 100$, while for
$c_{s*}\sim 0.7$ we already have that $n_s \simeq 0.95$ for this same range of $Q_*$
values.
The optimum range for $c_{s*}$ for this specific model is found to be $0.9 \lesssim
c_{s*} \lesssim 1$, for which the spectral tilt can be kept within the $95\%$ CL value
$n_s=0.9658\pm 0.0040$, obtained from the Planck TT,TE,EE+lowE+lensing+BK15+BAO+running data.} 
for $n_s$ the larger is $Q_*$ and the
smaller is $c_{s*}$, showing that the spectrum  is already outside the
two-sigma range in the red-tilted side of the observational values for
$n_s$. This indicates that for this form of a cubic dissipation
coefficient, the model is not expected to benefit from a DBIWI
realization, as already discussed before. Let us then next focus on
the linear dissipation  coefficient WI model of
Ref.~\cite{Bastero-Gil:2016qru}.  Some explicit examples of results
obtained in this case are given in Table~\ref{tab2} for the
quartic inflaton potential and with a linear in $T$ dissipation
coefficient. {}As it is clear, for $Q_*>10$, the tensor-to-scalar
ratio is smaller than $10^{-7}$, hence, the model is able to resolve
the aforementioned inconsistency in the cold DBI
inflation~\cite{Lidsey:2007gq} and discussed in the Introduction
section.

\begin{widetext}
\begin{center}
\begin{table}[!htb]
\caption{Numerical values of the parameters and the relevant
  cosmological quantities obtained for the case of a quartic inflaton
  potential ($n=2$) with a linear in $T$ dissipation coefficient
  ($c=1$, $p=0$) and for $c_{s*}=0.1$, when $G(Q_*) = 1$, and for
  $c_{s*}=0.3$, when $G(Q_*)$ is given by Eq. (\ref{Growingmode}).}
\begin{tabular}{c|c|c|c|c|c|c|c|c|c|c|c|c}
\hline\hline  $c_{s*}$ &$Q_*$ & $n_s$ & $r$ & $N_*$ & $|\Delta \phi|/M_{\rm Pl}$ & $C_\Upsilon$ &  $T_{\rm end}$ (GeV) & $V_0$ (GeV)$^4$& $\epsilon_{V_*}$ & $\eta_{V_*}$  & $V_*^{1/4}/M_{\rm Pl}$& $f_{0}$\\ 
\hline
 &1.03 & 0.9651 & $1.13\times 10^{-5}$ & 62.1 & 3.36 & 0.0067  & $2.67 \times 10^{13}$ & $3.94\times 10^{59}$ & 0.68 & 1.02 & $1.56\times 10^{-5}$ & $3.30 \times 10^{17}$ \\ 
0.1 & 10.25 & 0.9628 &  $3.00\times 10^{-7}$ & 61.0 & 1.42 & 0.023 & $1.08\times 10^{13}$ & $3.27 \times 10^{59}$ & 3.80& 5.71  & $6.31\times 10^{-6}$ & $2.14 \times 10^{18}$ \\ 
 & 102.51 & 0.9625 & $3.88\times 10^{-9}$ & 60.0 & 0.46 & 0.076  & $3.68 \times 10^{12}$ & $3.69\times 10^{59}$ & 35.55& 53.33 & $2.16\times 10^{-6}$ & $1.71 \times 10^{19}$ \\
\hline
 & 1.03 & 0.9672 & $3.33 \times 10^{-5}$ & 62.4  & 5.85 & 0.0088  & $3.27 \times 10^{13}$ & $1.27 \times 10^{59} $ & 0.23 &  0.34 & $1.91 \times 10^{-5}$ & $3.15 \times 10^{17}$\\ 
0.3 & 10.30  &  0.9667 & $5.70 \times 10^{-7}$  & 61.5  & 2.44 &  0.027 & $ 1.09 \times 10^{13} $ & $7.24 \times 10^{58} $ & $1.30$ & 1.95 & $6.39 \times 10^{-6}$ & $2.91 \times 10^{18}$\\ 
  & 103.16&  0.9684 & $3.49 \times 10^{-9}$ & 60.3 & 0.79 & 0.074 & $3.08\times 10^{12}$ & $3.40\times 10^{58}$ & 12.31 & 18.47 & $1.80 \times 10^{-6}$ & $4.73\times 10^{19}$\\ \hline \hline
\end{tabular}
\label{tab2}
\end{table}
\end{center}
\end{widetext}


\section{Conclusions}
\label{conclusions}

We studied the effects of a low sound speed  on the dynamics of
perturbations equations of WI inspired by string motivated models that
include relativistic D-brane motion. We numerically solved the
coupled inflaton and radiation field perturbation equations for the
first time in the case of a noncanonical kinetic term in DBIWI. We
found that a low sound speed is able to suppress the growing function
that always appears in the scalar power spectrum of WI,  whenever the
dissipation coefficient  exhibits an explicit dependence with the
temperature of the radiation bath. As a consequence of this
suppression effect seen in DBIWI, the restrictions for constructing a
WI realization in the strong dissipation regime $Q \gg 1$ are
considerably relaxed. We have also complemented these results with the
ones derived from the analytical expressions for the spectral tilt.
Based on our results, as the low sound speed will push WI models into
the strong dissipation regime, the severe inconsistency problems seen
in DBI cold inflation, e.g.,  due to an upper bound on the
tensor-to-scalar ratio arising from compactification constraints and a
lower bound from observations, can all be resolved due to large
dissipation. We have explicitly shown that among the most common
dissipation coefficients that have been derived from well-motivated
particle physics realizations and applied to the WI context, the
dissipation coefficient with a linear dependence on the temperature is
the case that can mostly benefit from a DBIWI realization.  Hence,
these type of models are able to soften all theoretical and
observational constraints, while predicting very small
tensor-to-scalar ratio and potentially significant non-Gaussianity,
making them falsifiable in near future. 
 
{}From a model-building perspective, the present work, along with the
previous results on swampland conjectures in WI~\cite{Kamali:2019xnt},
gives a strong hint that WI may consistently be embedded in string
theory utilizing the physics of the brane. Besides, phenomenologically
realizing WI in the strong dissipation regime is a significant step
towards achieving such goal. Therefore, the last step is building an
explicit model describing how D-brane may be able to dissipate its
energy into radiation field. This and other implications of our
results are certainly worthwhile to explore further in future
studies.

\section*{Acknowledgments}

R.O.R. is partially supported by research grants from Conselho
Nacional de Desenvolvimento Cient\'{\i}fico e Tecnol\'ogico (CNPq),
Grant No. 302545/2017-4, and Funda\c{c}\~ao Carlos Chagas Filho de
Amparo \`a Pesquisa do Estado do Rio de Janeiro (FAPERJ), Grant
No. E-26/202.892/2017.
 


\end{document}